\documentclass[aps,prl,preprint,superscriptaddress,preprintnumbers,nofootinbib,tightenlines,floatfix]{revtex4-1}

\usepackage{graphicx} % Include figure files
\usepackage{dcolumn}  % Align table columns on decimal point
\usepackage{bm}       % bold math
\usepackage{multirow}

\def\Dsp{D_{s}^{+}}
\def\Dsm{D_{s}^{-}}
\def\Ds{D_{s}}
\def\Dssp{D_{s}^{*+}}

\def\Dss{D_{s}^{*}}
\def\electron{e^{-}}
\def\positron{e^{+}}
\def\Dsstee{D_{s}^{*+} \to D_{s}^{+} e^{+} e^{-}}
\def\Dsstgamma{D_{s}^{*+} \to D_{s}^{+} \gamma}
\def\Dsstpi{D_{s}^{*+} \to D_{s}^{+} \pi^0}
\def\dsKKpi{K^+K^-\pi^+}
\def\dsKsK{K^0_SK^+}
\def\dspietaA{\eta\pi^+}

\def\dspietaprimeA{\eta'\pi^+}
\def\dspietaprimeB{\eta' \to \pi^+\pi^-\eta}

\def\dsKKpipizero{K^+K^-\pi^+\pi^0}
\def\dspipipi{\pi^+\pi^-\pi^+}

\def\dsKsKmpipi{K_{S}^0 K^- \pi^+ \pi^+}
\def\dspipizeroetaA{\eta\rho^+}

\def\dspietaprimerhoA{\eta'\pi^+}
\def\dspietaprimerhoB{\eta' \to \rho^0\gamma}
\def\invpb{\textrm{pb}^{-1}}
\def\Ree{R_{ee}}
\def\calB{{\cal B}}
\def\Eq#1{Eq.\ (\ref{#1})}

\def\epsiloneei{\epsilon_{ee}^i}
\def\epsilongammai{\epsilon_{\gamma}^i}
\def\ngammai{y_{\gamma}^i}
\def\yeei{n_{ee}^i}
\def\beei{b_{ee}^i}

\begin{document}

\title{\boldmath 
Observation of the Dalitz Decay $\Dsstee$
}

\preprint{CLNS 11/2074}  % the CLNS number
\preprint{CLEO 11-2 }    % the CLEO number

\author{D.~Cronin-Hennessy}
\author{J.~Hietala}
\affiliation{University of Minnesota, Minneapolis, Minnesota 55455, USA}
\author{S.~Dobbs}
\author{Z.~Metreveli}
\author{K.~K.~Seth}
\author{A.~Tomaradze}
\author{T.~Xiao}
\affiliation{Northwestern University, Evanston, Illinois 60208, USA}
\author{L.~Martin}
\author{A.~Powell}
\author{G.~Wilkinson}
\affiliation{University of Oxford, Oxford OX1 3RH, UK}
\author{H.~Mendez}
\affiliation{University of Puerto Rico, Mayaguez, Puerto Rico 00681}
\author{J.~Y.~Ge}
\author{D.~H.~Miller}
\author{I.~P.~J.~Shipsey}
\author{B.~Xin}
\affiliation{Purdue University, West Lafayette, Indiana 47907, USA}
\author{G.~S.~Adams}
\author{D.~Hu}
\author{B.~Moziak}
\author{J.~Napolitano}
\affiliation{Rensselaer Polytechnic Institute, Troy, New York 12180, USA}
\author{K.~M.~Ecklund}
\affiliation{Rice University, Houston, Texas 77005, USA}
\author{J.~Insler}
\author{H.~Muramatsu}
\author{C.~S.~Park}
\author{L.~J.~Pearson}
\author{E.~H.~Thorndike}
\affiliation{University of Rochester, Rochester, New York 14627, USA}
\author{S.~Ricciardi}
\affiliation{STFC Rutherford Appleton Laboratory, Chilton, Didcot, Oxfordshire, OX11 0QX, UK}
\author{C.~Thomas}
\affiliation{University of Oxford, Oxford OX1 3RH, UK}
\affiliation{STFC Rutherford Appleton Laboratory, Chilton, Didcot, Oxfordshire, OX11 0QX, UK}
\author{M.~Artuso}
\author{S.~Blusk}
\author{R.~Mountain}
\author{T.~Skwarnicki}
\author{S.~Stone}
\author{L.~M.~Zhang}
\affiliation{Syracuse University, Syracuse, New York 13244, USA}
\author{G.~Bonvicini}
\author{D.~Cinabro}
\author{A.~Lincoln}
\author{M.~J.~Smith}
\author{P.~Zhou}
\author{J.~Zhu}
\affiliation{Wayne State University, Detroit, Michigan 48202, USA}
\author{P.~Naik}
\author{J.~Rademacker}
\affiliation{University of Bristol, Bristol BS8 1TL, UK}
\author{D.~M.~Asner}
\altaffiliation[Now at: ]{Pacific Northwest National Laboratory, Richland, WA 99352}
\author{K.~W.~Edwards}
\author{K.~Randrianarivony}
\author{G.~Tatishvili}
\altaffiliation[Now at: ]{Pacific Northwest National Laboratory, Richland, WA 99352}
\affiliation{Carleton University, Ottawa, Ontario, Canada K1S 5B6}
\author{R.~A.~Briere}
\author{H.~Vogel}
\affiliation{Carnegie Mellon University, Pittsburgh, Pennsylvania 15213, USA}
\author{P.~U.~E.~Onyisi}
\author{J.~L.~Rosner}
\affiliation{University of Chicago, Chicago, Illinois 60637, USA}
\author{J.~P.~Alexander}
\author{D.~G.~Cassel}
\author{S.~Das}
\author{R.~Ehrlich}
\author{L.~Gibbons}
\author{S.~W.~Gray}
\author{D.~L.~Hartill}
\author{B.~K.~Heltsley}
\author{D.~L.~Kreinick}
\author{V.~E.~Kuznetsov}
\author{J.~R.~Patterson}
\author{D.~Peterson}
\author{D.~Riley}
\author{A.~Ryd}
\author{A.~J.~Sadoff}
\author{X.~Shi}
\author{W.~M.~Sun}
\affiliation{Cornell University, Ithaca, New York 14853, USA}
\author{J.~Yelton}
\affiliation{University of Florida, Gainesville, Florida 32611, USA}
\author{P.~Rubin}
\affiliation{George Mason University, Fairfax, Virginia 22030, USA}
\author{N.~Lowrey}
\author{S.~Mehrabyan}
\author{M.~Selen}
\author{J.~Wiss}
\affiliation{University of Illinois, Urbana-Champaign, Illinois 61801, USA}
\author{J.~Libby}
\affiliation{Indian Institute of Technology Madras, Chennai, Tamil Nadu 600036, India}
\author{M.~Kornicer}
\author{R.~E.~Mitchell}
\author{M.~R.~Shepherd}
\author{C.~M.~Tarbert}
\affiliation{Indiana University, Bloomington, Indiana 47405, USA }
\author{D.~Besson}
\affiliation{University of Kansas, Lawrence, Kansas 66045, USA}
\author{T.~K.~Pedlar}
\affiliation{Luther College, Decorah, Iowa 52101, USA}
\collaboration{CLEO Collaboration}
\noaffiliation

\date{\today}

\begin{abstract}
Using 586~$\invpb$ of $\positron\electron$ collision data acquired at $\sqrt{s}=4.170$~GeV with the CLEO-c detector at the Cornell Electron Storage Ring, we report the first observation of $\Dsstee$ with a significance of $5.3 \sigma$. The ratio of branching fractions $\calB(\Dsstee) / \calB(\Dsstgamma)$ is measured to be $[ 0.72^{+0.15}_{-0.13} (\textrm{stat}) \pm 0.10 (\textrm{syst}) ] \%$, which is consistent with theoretical expectations.
\end{abstract}


\maketitle

Dalitz decays \cite{DalitzDecay}, in which a virtual photon is internally converted to an $\positron\electron$ pair, have been observed in several vector-to-pseudoscalar decays of light mesons (\textit{e.g.}, $\omega\to\pi^0\positron\electron$, $\phi\to\pi^0\positron\electron$, and $\phi\to\eta\positron\electron$) \cite{pdg2010}. However, such decays have not been reported in electromagnetic decays of mesons containing charm or bottom quarks. This Letter reports the first observation of such a decay, $\Dsstee$, and a measurement of its branching fraction. Only two decay modes of the $\Dssp$ have been previously observed, the dominant $\Dsstgamma$ mode and the isospin-violating $\Dsstpi$~\cite{CLEO-DsstarDspi0} decay. Their branching fractions have been determined by the PDG~\cite{pdg2010} from measurements of the ratio $\calB(\Dsstpi) / \calB(\Dsstgamma)$ and the assumption that they are the only $\Dssp$ decay modes.

The expected $\Dssp$ Dalitz decay rate may be calculated by treating the photon from $\Dsstgamma$ as virtual and coupling it to an $\positron\electron$ pair. The $q^2$-derivative of the ratio 
\begin{equation}
\Ree\equiv{{\calB(\Dsstee)} \over {\calB(\Dsstgamma)}}
\end{equation}
can be written as \cite{Landsberg}
\begin{eqnarray}
\label{eq:dB}
{d\Ree \over dq^2}& =& {\alpha \over 3\pi q^2} \left|{f(q^2) \over f(0)}\right|^2 \left[{1-{4m_e^2 \over q^2}}\right]^\frac{1}{2}
\left[1+{2m_e^2 \over q^2}\right] \nonumber\\
&\times& \left[\left(1+{q^2 \over A}\right) - {4m_{\Dss}^2q^2 \over A^2}\right]^\frac{3}{2},
\end{eqnarray}
where $q$ is the four-momentum of the virtual photon, $m_x$ represents the mass of particle $x$, $A \equiv m_{\Dss}^2 - m_{\Ds}^2$, and $f(q^2)$ is the transition form factor for $\Dssp$ to $\Dsp$. Motivated by vector-meson dominance, we use $f(q^2)/f(0)=(1-q^2/m_{\phi}^2)^{-1}$. Integrating \Eq{eq:dB}, we predict $\Ree = 0.65\%$.

We use $\approx 5.6\times10^{5}$  $e^+e^-\to D_{s}^{\pm} D_{s}^{*\mp}$ events obtained from 586~$\invpb$ of $\positron\electron$ collision data with $\sqrt{s}$ near 4.170~GeV acquired by the CLEO-c detector at the Cornell Electron Storage Ring (CESR). The CLEO-c detector is equipped with a CsI(Tl) calorimeter~\cite{CLEOIIDetector} to detect photons and determine their directions and energies, and two concentric cylindrical wire drift chambers~\cite{Peterson2002142} to track the trajectory of charged particles. The tracking chambers operate in an axial 1~T magnetic field to provide momentum measurements. The beam pipe and the drift chambers present under 2\% of a radiation length of material, minimizing multiple scattering of charged particles and photon conversions. Charged hadron identification is achieved using energy loss ($dE/dx$) in the drift chambers and Cherenkov radiation in the RICH detector~\cite{DHad281,Artuso:2005dc}.

The default Kalman filter track reconstruction used to process CLEO data includes corrections for $dE/dx$ and multiple scattering in the beam pipe and detector material, assuming each track has the mass of a pion, kaon, and proton. The $e^\pm$ tracks in this analysis are rather soft, with energies below 150~MeV, where $dE/dx$ is very different from that of any of those three mass hypotheses. Therefore, to improve sensitivity, we reprocess events containing at least one exclusively reconstructed $\Dsp$ candidate, adding an $e^\pm$ mass hypothesis for each charged particle. Details of this analysis appear in Ref.~\cite{souvik2011}. CLEO has previously observed two other Dalitz decays, $\eta\to e^+e^-\gamma$ \cite{cleo:etadalitz} and $\eta^\prime\to e^+e^-\rho^0$ \cite{cleo:eta'dalitz}, in both of which the $e^{\pm}$ were substantially more energetic and did not need reprocessing.

We reconstruct $\Dsstee$ candidates using the nine distinct hadronic decay modes of the $\Dsp$ listed in Table \ref{tab:Unblinded_K}. Charge conjugate modes are also included. Candidates for $K_S^0$ and $\eta$ are reconstructed through their decays to $\pi^+\pi^-$ and $\gamma\gamma$, respectively. Measurement of $\Ree$ instead of the absolute branching fraction $\calB(\Dsstee)$ bypasses any need for estimating the total number of $\Dssp$ produced and minimizes systematic uncertainties stemming from reconstruction of the $\Dsp$.

We follow a blind analysis procedure to avoid bias. Selection criteria are optimized individually in each of the $\Dsp$ decay modes for maximum signal significance using Monte Carlo simulated samples of signal and background processes. The decay chain of signal events $\positron\electron \to \Dssp\Dsm; \Dsstee$ are simulated with full angular correlations. Simulated samples of all $\positron\electron \to q\bar{q}$ processes at 4.170~GeV where $q=u,d,s$, or $c$ are used for background. The reconstructed $\positron\electron$ tracks are required to pass within 5~cm of the interaction point in the direction parallel to the beam-axis and within 5~mm of the beam-axis in the transverse directions. The $dE/dx$ of each $e^\pm$ candidate is required to be within $3\sigma$ of that expected for electrons. All charged pions and kaons in the $\Dsp$ decay chain are identified as such using a combination of $dE/dx$ and RICH information as described in Ref.~\cite{:2008cqa}. We require the reconstructed $\Dsp$ mass $M_{\Ds}$ to be within a mode-dependent region around the known $\Dsp$ mass~\cite{pdg2010} consistent with the resolution of the detector.  We define the beam-constrained mass of the $\Dssp$ by $M_{\mathrm{BC}} \equiv \sqrt{ E_{\Dss}^2 - \mathbf{p}_{\Dss}^2 }$, where $E_{\Dss}$ is the energy of the $\Dssp$ calculated from the beam energy and $\mathbf{p}_{\Dss}$ is the three-momentum of the $\Dssp$ inferred from its decay daughters' measured momenta. We select events with $M_{\mathrm{BC}}$ and  $\delta M \equiv M_{\Dss} - M_{\Ds}$ consistent with the known $\Dssp$ and $\Dsp$ masses~\cite{pdg2010}.

A significant background to the observation of $\Dsstee$ arises from $\Dsstgamma$ events where the $\gamma$ converts into an $\positron\electron$ pair in the material of the beam-pipe or drift chambers. We reject much of this background using the following criteria for the $e^\pm$ tracks. We define the $d_0$ of a track as the distance of closest approach of the track to the beam axis. Its sign depends on the charge of the track and whether the origin of the $x-y$ plane falls within the circle of the track in that plane. We require the difference between the $d_0$ of the $\positron$ and $\electron$ tracks, $\Delta d_0 = d_0^- - d_0^+$, to exceed $-5$~mm. Denoting each electron track's azimuthal angle measured at the point of closest approach to the beam axis by $\phi_0$, we also require $\Delta \phi_0 = \phi_0^- - \phi_0^+ < 0.12$.

These selection criteria are applied on simulated samples of our signal to obtain the selection efficiencies for signal events $\epsiloneei$, where $i$ stands for one of the nine decay modes of the $\Dsp$ used in this analysis. These criteria are applied to data in order to obtain the yields of events $\yeei$. These numbers are presented in Table \ref{tab:Unblinded_K} for each decay mode of the $\Dsp$.

\begin{figure}[t]
\centering
\includegraphics[width=0.60\linewidth]{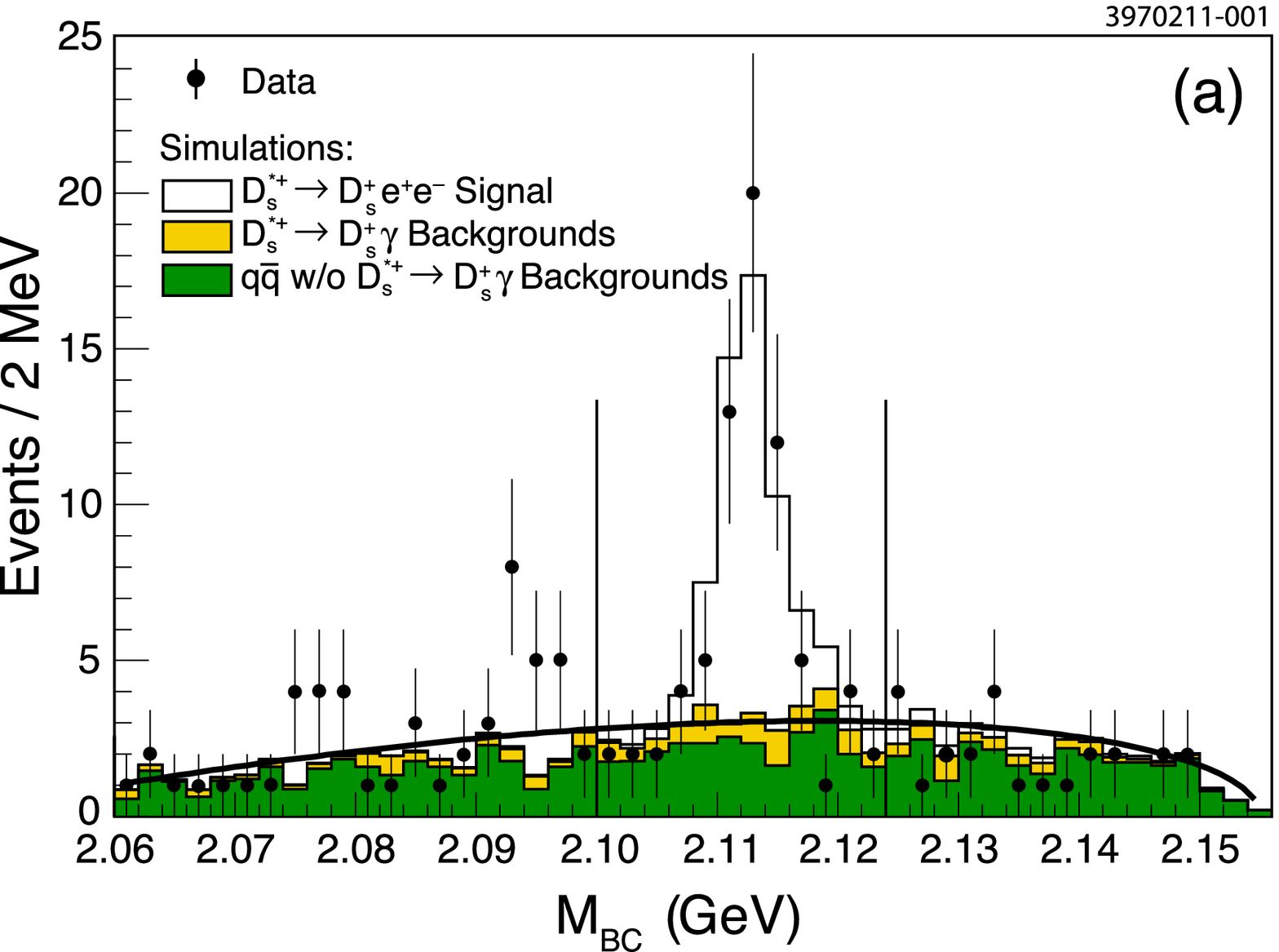}
\includegraphics[width=0.60\linewidth]{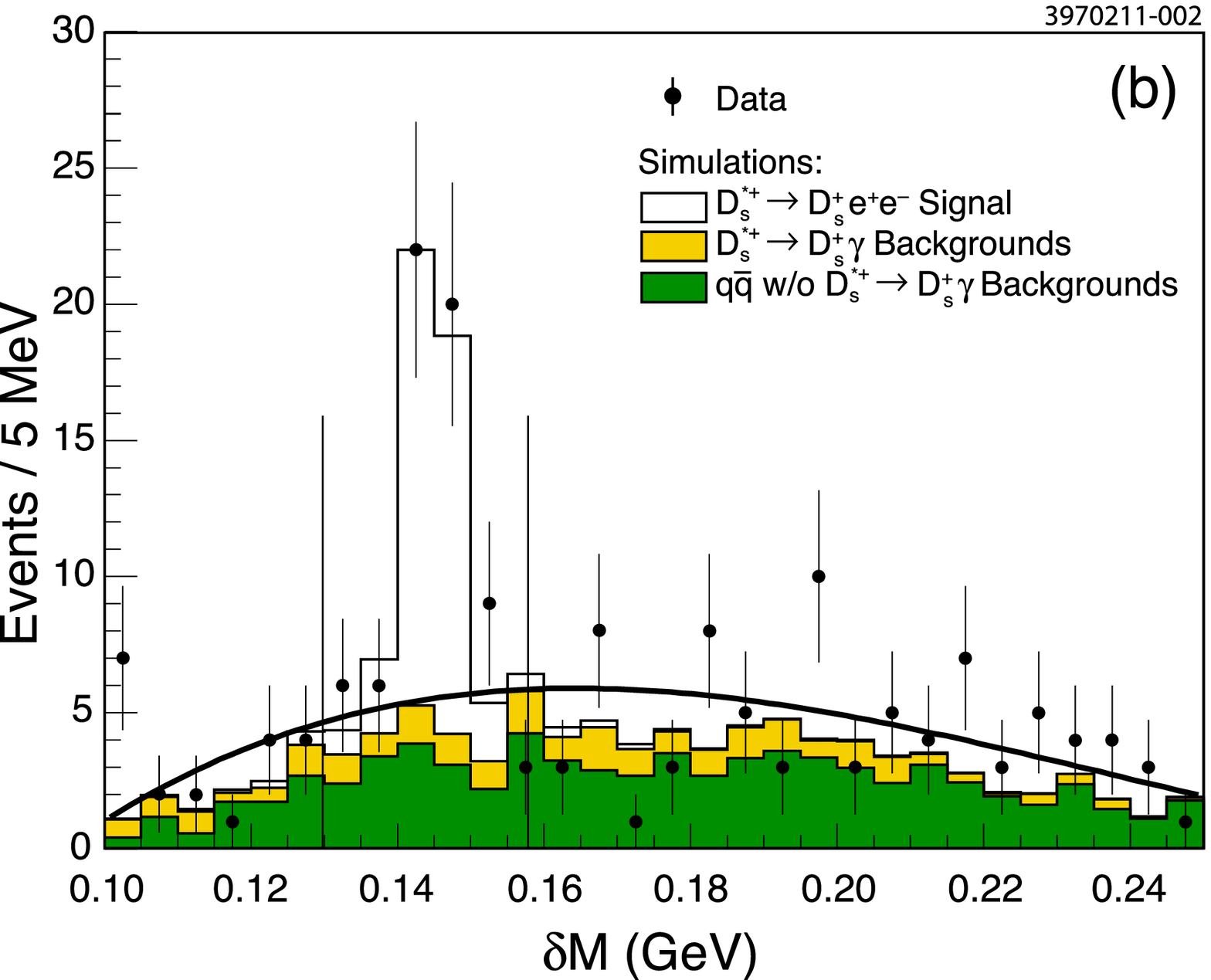}
\caption{Distributions of (a) $M_{\mathrm{BC}}$ and (b) $\delta M$ in data and simulated samples summed over all nine $\Dsp$ decay modes used in this analysis. In each figure, the points with error bars are data, and the unshaded histogram is the simulated $\Dsstee$ signal. Background events in the upper shaded histogram (yellow online) are from simulated $\Dsstgamma$ decays. Background events in the lower shaded histogram (green online) are from simulated $q\bar{q}$ events that do not include $\Dsstgamma$ or $\Dsstee$. The curves are the fits of data to background shapes described in the text. The regions 2.100 to 2.124 GeV in (a), and 0.1298 to 0.1578 GeV in (b) are avoided in the shape fits to prevent contamination of the background estimates with signal events.\label{fig:Sum_Unblind}}
\end{figure}

Having established selection criteria using simulations, the background $b_{ee}^i$ in the signal region for each mode $i$ is estimated from the fit of an $M_{BC}$ background function to the data in the $M_{BC}$ sidebands for that mode. The shape of the $M_{BC}$ function is fixed and is common to all modes. The function incorporates the kinematic limit and its parameters are determined from simulations. This shape is illustrated in Fig. 1(a). A similar estimate of $b_{ee}^i$ obtained from the $\delta M$ distribution and the difference between this estimate and the $M_{BC}$ estimate is taken as the systematic uncertainty inherent in the procedure. The estimated background for each mode is presented in Table~\ref{tab:Unblinded_K} as $\beei$. 

We calculate the signal significance, expressed in number of Gaussian standard deviations, from the Poisson probability for the estimated background to fluctuate up to the observed signal yield or higher. The uncertainties in the background, both statistical and systematic, are modeled as Gaussian distributions. The combined signal significance for all modes is $5.3\sigma$. The most significant individual mode is $\dsKKpi$ at $5.0\sigma$.

\begin{figure}[htb]
\centering
\includegraphics[width=\linewidth]{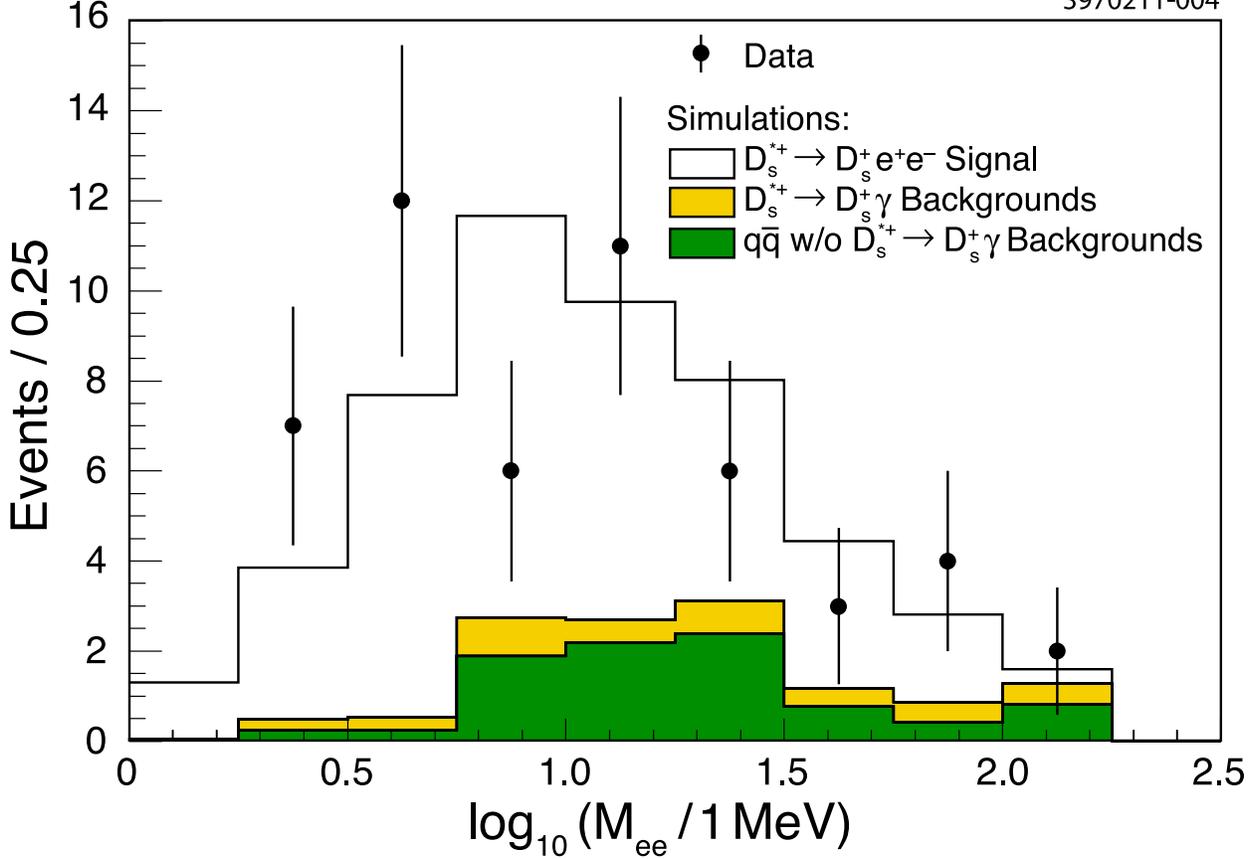}
\caption{Distribution of $M_{ee}$ in simulated events within the signal region overlaid with the 51 events observed in data. The interpretations of the various simulation histograms are identical to those in Fig.~\ref{fig:Sum_Unblind}.}
\label{fig:mee_Shape}
\end{figure}

The distribution of the $\positron\electron$ invariant mass $M_{ee}$ for the 51 observed events in the signal region is compared to that expected in our simulations and presented in Fig.~\ref{fig:mee_Shape}. The distributions are found to be in good agreement, with a Kolmogorov-Smirnov probability of 0.86 for the events to share the same parent distribution.

\begin{table*}[htb]
\centering
\caption{
The yields $\yeei$, estimated backgrounds $\beei$, background-subtracted yields $\ngammai$, ratios of detection efficiencies $\xi^i$, and values $\Ree^i$ of $\Ree$ for each $\Dsp$ decay mode $i$. The uncertainties given for $\xi^i \equiv \epsilon_\gamma^i/\epsilon_{ee}^i$ are statistical only. The values of $\yeei$ and $\beei$ are summed over all modes, while the ratio of branching fractions $\Ree$ is computed using \Eq{eq:Ree}.}
\setlength{\tabcolsep}{0.5pc}
\begin{tabular}{cccccc}
\hline
\hline\noalign{\smallskip}
$i$ & $\yeei$ & $\beei$ & $\ngammai$ & $\xi^i$ & $\Ree^i$(\%) \\
\hline\noalign{\smallskip}
$\dsKKpi$ & 14 & $1.05^{+0.42}_{-0.33} \pm 0.79$ & $9114 \pm 110 \pm 201$ & $4.65 \pm 0.12$ & $0.66^{+0.21}_{-0.18}$ \\
\noalign{\smallskip}
$\dsKKpipizero$ & 6 & $1.70^{+0.52}_{-0.43} \pm 0.56$ & $3592 \pm 118 \pm 72$ & $4.80 \pm 0.21$ & $0.58^{+0.38}_{-0.29}$ \\
\noalign{\smallskip}
$\dsKsK$ & 1 & $0.85^{+0.50}_{-0.36} \pm 0.74$ & 1902 $\pm$ 57 $\pm$ 45 & $4.31 \pm 0.13$ & $0.03^{+0.33}_{-0.18}$ \\
\noalign{\smallskip}
$\dsKsKmpipi$ & 4 & $1.58^{+0.59}_{-0.47} \pm 0.40$ & 1570 $\pm$ 74 $\pm$ 13 & $5.38 \pm 0.20$ & $0.83^{+0.83}_{-0.60}$ \\
\noalign{\smallskip}
$\dspipipi$ & 7 & $1.57^{+0.50}_{-0.41} \pm 0.59$ & $2745 \pm 93 \pm 52$ & $4.62 \pm 0.10$ & $0.91^{+0.51}_{-0.40}$ \\
\noalign{\smallskip}
$\dspietaA$ & 4 & $1.40^{+0.82}_{-0.59} \pm 0.49$ & $1037 \pm 46 \pm 37$ & $3.87 \pm 0.10$ & $0.97^{+0.93}_{-0.67}$ \\
\noalign{\smallskip}
$\dspipizeroetaA$ & 7 & $2.62^{+0.63}_{-0.54} \pm 0.23$ & $3170 \pm 161 \pm 313$ & $5.82 \pm 0.24$ & $0.80^{+0.56}_{-0.44}$ \\
\noalign{\smallskip}
$\dspietaprimeA; \dspietaprimeB$ & 4 & $0.00^{+0.72}_{-0.00} \pm 0.00$ & $691 \pm 34 \pm 40$ & $3.96 \pm 0.12$ & $2.30^{+1.50}_{-0.97}$ \\
\noalign{\smallskip}
$\dspietaprimerhoA; \dspietaprimerhoB$ & 4 & $1.84^{+0.54}_{-0.45} \pm 0.25$ & $1531 \pm 80 \pm 122$ & $4.97 \pm 0.14$ & $0.70^{+0.78}_{-0.57}$ \\
\noalign{\smallskip}
\hline
\noalign{\smallskip}
All Modes & 51 & $12.61^{+1.78}_{-1.29} \pm 4.05$ &  &  & $0.72^{+0.15}_{-0.13}$ \\
\noalign{\smallskip}
\hline
\hline
\end{tabular}
\label{tab:Unblinded_K}
\end{table*}

\begin{figure}[t]
\centering
\includegraphics[width=\linewidth]{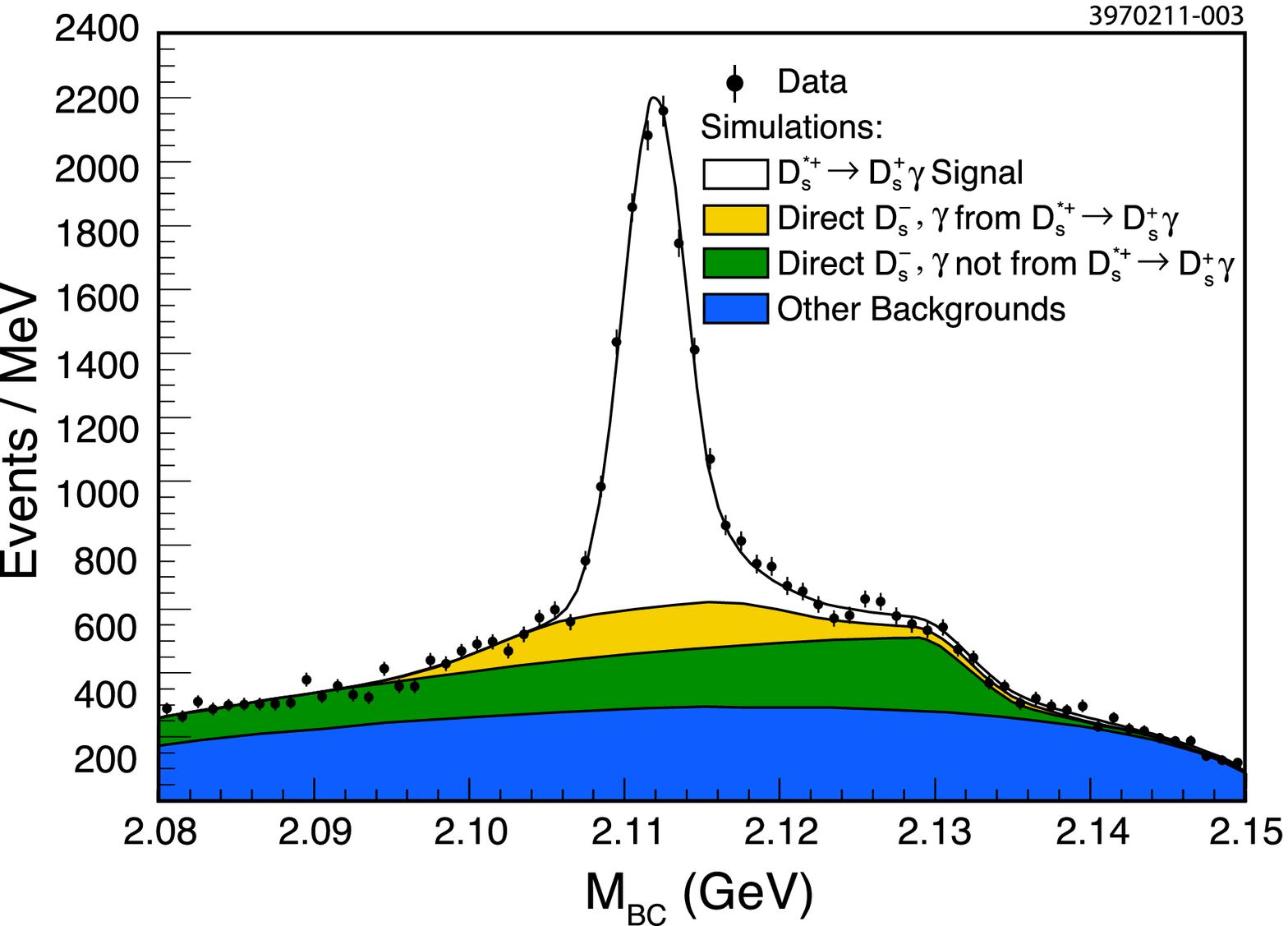}
\caption{Distribution of $M_{\mathrm{BC}}$ of $\Dsstgamma$ events where $\Dsp \to \dsKKpi$. The points with error bars are data, the unshaded region is the $\Dsstgamma$ signal. Background events in the highest shaded region (yellow online) are events with a direct $\Dsm$ paired with a photon from the $\Dsstgamma$ decay. Background events in the middle shaded region (green online) are events with a direct $\Dsm$ paired with a photon that did not come from $\Dsstgamma$. Background events in the lowest shaded region (blue online) are from all other sources. \label{fig:KKpi_DsGamma_Data_MBC}}
\end{figure}

The criteria for selecting $\Dsstgamma$ events follow those for $\Dsstee$ as closely as possible. Instead of an $\positron\electron$ pair, a photon candidate in the kinematically allowed energy range is required and its electromagnetic shower in required to have a lateral spread narrower than 99\% of true photons. The allowed window for $\delta M$ is broadened to account for worse resolution in calorimetric photon energy as compared to tracking-based $\positron\electron$ measurements. For similar reasons, the mass window restriction on $M_{BC}$ is removed and the signal yield is extracted from a fit using background and signal shapes determined from simulations, as depicted in Fig.~\ref{fig:KKpi_DsGamma_Data_MBC} for the $\dsKKpi$ mode. Application of these criteria to simulated samples of $\Dsstgamma$ gives us the efficiencies $\epsilongammai$. Application to data gives us the background-subtracted signal yields $\ngammai$ as listed in Table~\ref{tab:Unblinded_K}.

The ratio $\xi^i \equiv \epsilon_\gamma^i/\epsilon_{ee}^i$ of efficiencies for $\Dsstgamma$ to $\Dsstee$, given in Table \ref{tab:Unblinded_K}, cluster around 4.8. The $\dsKKpi$ mode stands out due to its large $\Dsp$ branching fraction, presence of two charged kaons and the absence of photons in the final state. No other mode or combination of modes is predicted to have more significance for an observation of $\Dsstee$. For $\Dsp\to\dsKKpi$ (all other modes combined), we expect a yield of $\sim$15 (36) events, about 1 (13) of which are background. After unblinding the data as shown in Table~\ref{tab:Unblinded_K}, we find yields $\yeei$ quite close to those expected from our predicted $R_{ee}$ and Monte Carlo simulations. Yields from individual non-$\dsKKpi$ modes range from 1 to 7, consistent with expectations.

For each of the $\Dsp$ decay modes $i$, we calculate the value of $R_{ee}^i = ( \yeei - \beei )/( \ngammai/\xi^i )$ as listed in Table~\ref{tab:Unblinded_K}. Note that the values of $\xi^i$ are all equal ($\approx 4.8$) to within $\pm20\%$ because the dominant difference in efficiency is due to the photon versus $\positron\electron$ selection criteria, which are identical for all modes. The $\xi^i$ vary somewhat with $i$ due to the broadened $\delta M$ windows and signal shape $M_{BC}$ fit (instead of a fixed window) for the radiative $\Dssp$ modes relative to those of the Dalitz decays. Hence most systematic uncertainties in $\xi^i$ due to $D_s^+$ selection cancel; {\it e.g.} those due to track-finding, particle identification, and selection of photons, $\pi^0$, $K_S^0$ decays, {\it etc.} In order to avoid a bias due to Poisson fluctuations when $\yeei$ is small and to preserve the cancellation of systematic uncertainties inherent in the use of $\xi^i$, we calculate $R_{ee}$ by weighting each $R_{ee}^i$ by the \textit{expected} number of Dalitz decays, which is proportional to $\ngammai/\xi^i$. This gives
\begin{equation}
R_{ee} ={ {\sum_i ( \ngammai/\xi^i ) R_{ee}^i}\over{\sum_i \ngammai/\xi^i }}
= { { \sum_i \yeei - \beei }\over{\sum_i \ngammai/\xi^i} }.
\label{eq:Ree}
\end{equation}

We consider systematic uncertainties of the signal yields and efficiency ratios in our measurement of $\Ree$. The systematic uncertainty in $\beei$ contributes a fractional uncertainty of 10.6\%. Systematic uncertainties in $\ngammai$ and statistical uncertainties in $\xi^i$ contribute a total fractional systematic uncertainty of 2.0\%. Two systematic uncertainties remain; first, an uncertainty due to the different selection criteria on $M_{BC}$ and $\delta M$, as described above, which is estimated to be 4.1\%; second, the uncertainty in the ratio of reconstructing an $\positron\electron$ pair to that of a $\gamma$. The uncertainty in this ratio is estimated by studying the decay $\psi(2S) \to J/\psi\ \pi^0\pi^0$. The ratio between the number of events where one of the $\pi^0$ decays to $\gamma\positron\electron$ and the number of events where both $\pi^0$ decay to $\gamma\gamma$ must be equal to twice the ratio of branching fractions $R_{ee}^{\pi^0}\equiv{{\calB(\pi^0\to\gamma\positron\electron)}/{ \calB(\pi^0\to\gamma\gamma)}}$. Using this relationship, we measure $R^{\pi^0}_{ee}=(1.235\pm0.051)\%$ in a manner similar to $\Ree$ by reconstructing the $J/\psi\ \pi^0\pi^0$ through these two decay modes of the $\pi^0$. We restrict the energies of the $e^{\pm}$ to the range 20 to 144~MeV (the mass difference $m_{\Dss}-m_{\Ds}$). Compared to the PDG value of $(1.188\pm0.035)\%$ \cite{pdg2010}, we estimate the fractional systematic uncertainty on the ratio of efficiencies to be 6.5\%. These fractional systematic uncertainties are combined in quadrature to yield the final result for the ratio $\Ree$:
\begin{equation}
\label{eq:Result_final}
\Ree = [0.72^{+0.15}_{-0.13} (\textrm{stat}) \pm 0.10 (\textrm{syst})] \%.
\end{equation}
We use this result to re-evaluate the absolute branching fractions of the $\Dssp$ meson as presented in Table~\ref{tab:ReEval}.

In summary, we report the first observation of a third decay mode of the $\Dssp$, the $\Dsstee$. We observe 51 candidate events in our signal region with an expected background of 12.6~events. The signal significance is $5.3\sigma$. The ratio of branching fractions $\Ree$ is measured as presented in \Eq{eq:Result_final} and found to be consistent with our theoretical prediction. This implies that existing estimates of $\Dssp$ branching fractions should be revised.

\begin{table}[t]
\centering
\caption{Branching fractions in percent for the known decays of the $\Dssp$ meson from PDG 2010~\cite{pdg2010}, which assumed $B(\Dsstee)=0$, and our re-evaluation using the value of $\Ree$ reported in this Letter. \label{tab:ReEval}}
\setlength{\tabcolsep}{0.90pc}
\begin{tabular}{lcc}
\hline\hline
\rule[10pt]{-1mm}{0mm}
Decay Mode & PDG 2010 & This Analysis \\
\hline
\rule[10pt]{-1mm}{0mm}
$\Dsstgamma$ & 94.2$\pm$0.7 & 93.5$\pm$0.5$\pm$0.5 \\[1pt]
$\Dsstpi$    & 5.8$\pm$0.7 & 5.8$\pm$0.4$\pm$0.5 \\[1pt]
$\Dsstee$~~~ & 0   & $0.67^{+0.14}_{-0.12} \pm 0.09$ \\[1pt]
\hline\hline
\end{tabular}
\end{table}

\begin{acknowledgments}
We gratefully acknowledge the effort of the CESR staff in providing us with excellent luminosity and running conditions. This work was supported by the A.P.~Sloan Foundation, the National Science Foundation, the U.S. Department of Energy, the Natural Sciences and Engineering Research Council of Canada, and the U.K. Science and Technology Facilities Council. 
\end{acknowledgments}

\bibliography{Dsstee_Paper}

\end{document}